\documentclass[sn-mathphys-num,iicol]{sn-jnl}% Math and Physical Sciences Numbered Reference Style 
\usepackage{graphicx}%
\usepackage{multirow}%
\usepackage{amsmath,amssymb,amsfonts}%
\usepackage{amsthm}%
\usepackage{mathrsfs}%
\usepackage[title]{appendix}%
\usepackage{xcolor}%
\usepackage{textcomp}%
\usepackage{manyfoot}%
\usepackage{booktabs}%
\usepackage{algorithm}%
\theoremstyle{thmstyleone}%
\theoremstyle{thmstyletwo}%

\theoremstyle{thmstylethree}%

\begin{document}

%\title[Red giant branch stars as probes of metallicities]{Red giant branch stars as probes of metallicity evolution in the galaxies}

\title[Red giant branch stars as probes of metallicities]{Metallicity relations in LMC and SMC from the slope of Red Giant Branch stars in globular clusters.}

\author*[1]{\fnm{Saurabh} \sur{Sharma}}\email{saurabh@aries.res.in}

\author[2]{\fnm{Jura} \sur{Borissova}}\email{jura.borissova@uv.cl}

%\author[2]{\fnm{Radostin} %\sur{Kurtev}}\email{radostin.kurtev@uv.cl}

\affil*[1]{\orgdiv{Aryabhatta Research Institute of Observational Sciences (ARIES), Manora Peak, Nainital, 263 001, India}}

\affil[2]{\orgdiv{Instituto de Física y Astronomía, Universidad de Valparaíso, Ave. Gran Bretaña 1111, Valparaíso, Chile; Millennium Institute of Astrophysics, MAS, Chile}}

%\affil[3]{\orgdiv{Department}, \orgname{Organization}, \orgaddress{\street{Street}, \city{City}, \postcode{610101}, \state{State}, \country{Country}}}


\abstract{Red giants are an excellent tool for probing the 
history of star formation and subsequent metallicity evolution in the galaxies. The well-defined red giant branch (RGB) stars of the globular clusters can be used to determine their slopes and to calibrate the RGB slope parameters, age, and metallicity relations.
We obtained deep near-IR $JHK$ stellar photometry of 23 LMC/SMC globular clusters.
The cluster sample covers a wide range in metallicities $(-1.76<{\rm [Fe/H]}<-0.32)$ and ages $(0.6\,{\rm Gyr}<t<14\,{\rm Gyr})$. 
The slope of the RGBs of each cluster was calculated and used to derive the relations between slope-age-metallicity. 
 We have found that the RGB slope do not shows any statistically significant age dependence. 
%We have found that the oldest  ($>$ 10 Gyr) clusters have larger RGB slope values than young ($<$3 Gyr) ones.
The young and old clusters are found to be distributed differently in RGB slope-metallicity space, and the younger populations show a slightly less steep RGB slope dependence than the whole cluster sample.  The population of the younger clusters shows a negative slope, whereas the older clusters show a positive slope.  
}

   \keywords{Hertzsprung-Russel (HR) diagram -- globular clusters: general --
                galaxies: abundances -- galaxies: individual (LMC, SMC) -- infrared: stars
                }


\maketitle

\section{Introduction}\label{sec1}

Red giant branch (RGB) stars are among the brightest red stars in stellar
systems older than a Gyr. They appear in essentially all galaxies.
Therefore, red giants are an excellent tool for probing the parameters of
old populations and the history of star formation in galaxies.
Globular clusters, on the other hand, are ideal sites for calibrating the RGB parameters.
Since \cite{dacosta1} explored the possibility of using the position of the RGB as a metallicity indicator, many significant advances have been made in improving astronomical instrumentation and developing the corresponding theory. 
The infrared (IR) wavelength regime became accessible, and it is particularly compelling for such
studies, in comparison with the optical one, because of the potential to
probe the stellar populations of systems with high intrinsic extinction.
Aside from minimizing reddening effects, the near-IR spectral region
``straightens" the RGB, converting it from a hyperbola (as seen in
\cite{saviane}) to a straight line (\cite{ivanov2002}). This implies that
the depth of the photometry is not as crucial
in the near-IR as in the optical to establish a slope.  Red giants
are brighter in the IR, and the RGB slope is independent of distance
and reddening. Also, the IR can help break the age-metallicity degeneracy
that plagues the optical.

\cite{kuchinski} and  \cite{miniti} demonstrated both empirically and
theoretically that the slope of the RGB in a
$K$ vs. $(J-K)$ color-magnitude diagram (CMD) is sensitive to the metallicity of
the population. They investigated this correlation for metal-rich globular
clusters and derived a linear relation between $\rm [Fe/H]$ and RGB slope.
\cite{ivanov2000} extended this relation to metallicities
as low as $\rm [Fe/H]=-2.1$. \cite{ferraro2000} used high-quality data 
for ten Galactic globular clusters and
derived a similar relation. \cite{ivanov2002}
tabulated the relation in [$M_K$,$(J-K_S)$, $\rm [Fe/H]$] space based on
a large sample of 2MASS photometry of Milky Way globular clusters. 
%The RGBs were fitted by straight lines and produced new calibrations of the
%RGB slope, tip, and zero point are functions of abundance.
They fitted straight lines to the RGBs stars and produced new calibrations, where, 
RGB slope, tip, and zero point are the functions of abundances.
\cite{tiede}, using 4 Galactic open clusters, derived
the slope of the RGB - metallicity relation for a younger population, and
\cite{kyeong} re-derived the calibration, adding 10
new Galactic open clusters. Later on, \cite{tiede2002}
re-examined all existing data used to establish the calibration between
the slope of the upper giant branch in a $K$ vs. $J-K$ 
CMD  and the metallicity of the population of stars. They found two
very clear linear trends, one for Galactic open clusters spanning $-0.6
< {\rm [Fe/H]} < +0.1$ and another for Galactic globular clusters spanning
$-2.2 < {\rm [Fe/H]} < -0.4$.
\cite{valenti} presented new high\--quality near\--IR photometry
of 10 Galactic globular clusters spanning a wide metallicity range
($-$2.12 ${\leq}$ [Fe/H] ${\leq}$ $-$0.49). Five of their clusters belong to the
Halo and Five to the Bulge.  They constructed CMDs
in various planes (${\rm (K,J-K)}, {\rm (K,V-K)}, {\rm (H,J-H)}$, and
${\rm (H,V-H)}$) and measured a set of photometric indices
(colors, magnitudes, and slopes) describing the location and the
morphology of the RGB. They combined this new data set with those already published by
\cite{ferraro2000} and \cite{valentia}, and presented an updated calibration
of the various RGB indices in the 2MASS photometric system regarding the cluster metallicity. Finally, \cite{Kuinskas2008}, using a sample of 
intermediate-age (1-8 Gyr) clusters in the Large and Small Magellanic Clouds, 
found systematic differences between the RGB slope vs. metallicity relation 
derived from their sample and that of \cite{valentia}. They conclude  
that this discrepancy cannot be explained in terms of differences in cluster ages, and
its origin needs further investigation.
 
We attempt here to improve the relationship between RGB slope, metallicity 
and age by adding 23 young, intermediate, and old 
LMC and SMC clusters.

\begin{figure*}
\centering
\includegraphics[height=18cm,width=14cm,angle=0]{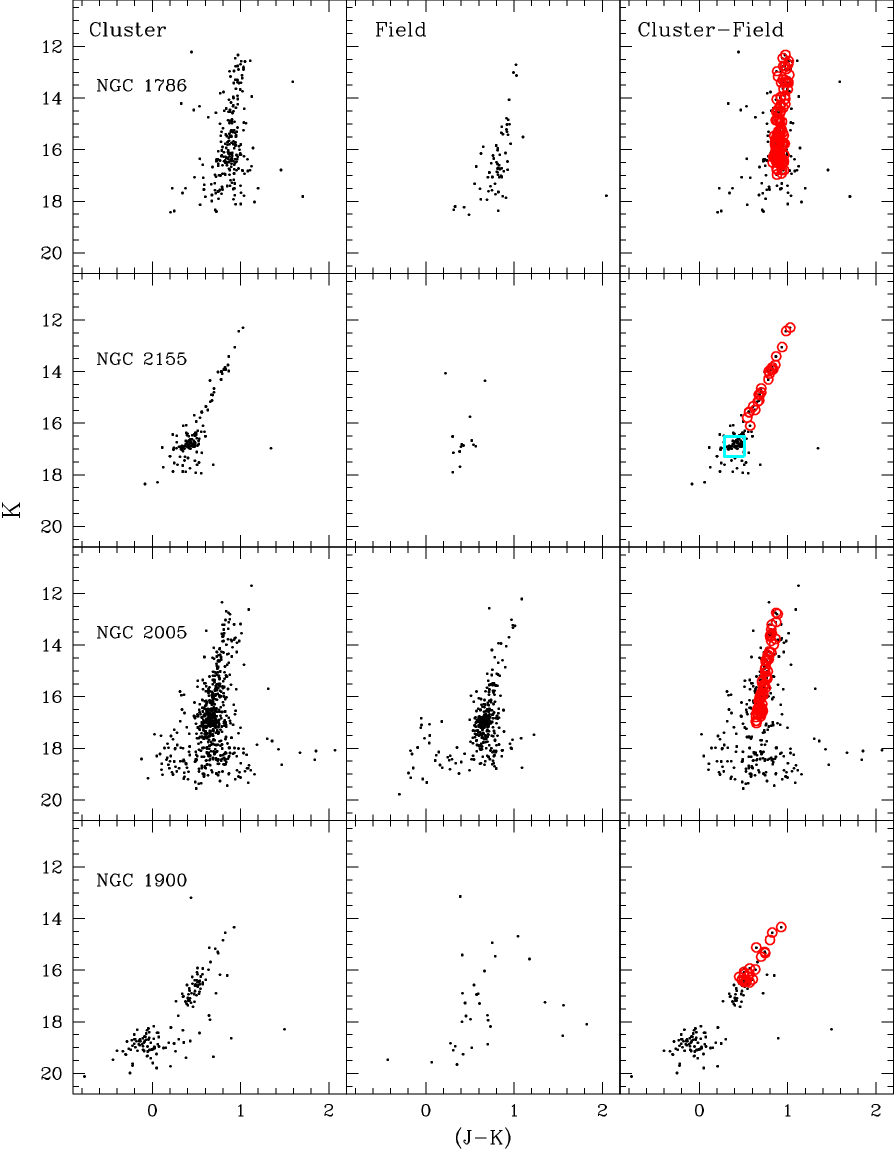}
\caption{\label{cmd} $(J-K), K$ CMDs for cluster region stars, field region stars, and statistically cleaned cluster member stars. The probable cluster member stars, used to calculate the slope of RGB, are shown in red circles. The location of the horizontal branch for the NGC 2155 cluster is shown with a cyan square box.
%The cluster RCs are denoted by square box.
}
\end{figure*}

\begin{table*}
%\begin{minipage}{0.7\textwidth}
%\centering
%\tiny
\begin{center}
\caption{\label{param} Cluster ages, metallicities, and RGB slope values.}
\begin{tabular}{@{}llcccccclll@{}}
\hline
    Cluster	& $\alpha_{(2000)}$& $\delta_{(2000)}$& $ [Fe/H] $&  $\sigma_{[Fe/H]}$&Slope$_{RGB}$ & $\sigma_{Slope}$& Log Age &$\sigma_{Age}$	&Ref & Ref \\
                & ($^\circ$) & ($^\circ$)  &  (dex)  &  &        &     & & &Age&  [Fe/H]\\
\hline
 ESO 121-3$^*$  &	90.5083	&	-60.5230   & -0.97&0.08  & -0.071  & 0.021   & 9.417    & 0.2      &  1        & 5  \\
Hodge 2$^*$	    &	79.4542	&	-69.6440    & -0.49&0.07  & -0.104  & 0.011   & 9.23     & 0.16     &  2        & 6  \\
Hodge 3	        &	83.3321	&	-68.1520    & -0.32&0.05  & -0.115  & 0.006   & 8.9      & 0.1      &  3        & 7  \\
Kron 28	        &	12.9148	&	-71.9991    & -0.94&0.04  & -0.087  & 0.005   & 9.57     & 0.04     &  4        & 2  \\
Kron 44  	    &	15.5264	&	-73.9230    & -0.78&0.07  & -0.09   & 0.005   & 9.57     & 0.08     &  4        & 2  \\
Lindsay 1$^*$   &	0.9750	&	-73.4720    & -0.98&0.07  & -0.095  & 0.004   & 9.86     & 0.216    &  2        & 4  \\
Lindsay 113$^*$	&	27.3750	&	-73.7280    & -1.12&0.05  & -0.074  & 0.005   & 9.56     & 0.195    &  2        & 8  \\
Lindsay 17	    &	8.9238	&	-73.5976    & -0.84&0.03  & -0.095  & 0.009   & 9.72     & 0.6      &  5        & 9  \\
Lindsay 7	    &	6.1798	&	-73.7533    & -0.76&0.05  & -0.09   & 0.006   & 9.3      & 0.27     &  5        & 9  \\
NGC 152  	    &	8.2365	&	-73.1157    & -0.72&0.06  & -0.09   & 0.005   & 9.28     & 0.07     &  2        & 2  \\
NGC 1754$^*$ 	&	73.5708	&	-70.4410    & -1.45&0.07  & -0.08   & 0.004   & 10.0     & 0.1      &  3        & 3  \\
NGC 1783	    &	74.7858	&	-65.9877    & -0.54&0.008 & -0.104  & 0.003   & 9.29     & 0.09     &  2        & 4  \\
NGC 1786$^*$	&	74.7833	&	-67.7450    & -1.76&0.3   & -0.12   & 0.003   & 10.06    & 0.06     &  3        & 3  \\
NGC 1795$^*$	&	74.9417	&	-69.8010    & -0.35&0.04  & -0.107  & 0.008   & 9.27     & 0.1      &  2        & 10 \\
NGC 1835$^*$	&	76.2792	&	-69.4070    & -1.69&0.05  & -0.091  & 0.006   & 10.15    & 0.05     &  3        & 3  \\
NGC 1846$^*$	&	76.8958	&	-67.4610    & -0.49&0.08  & -0.106  & 0.001   & 9.29     & 0.08     &  2        & 4  \\
NGC 1898$^*$	&	79.1708	&	-69.6570    & -1.15&0.05  & -0.098  & 0.005   & 10.07    & 0.08     &  2        & 3  \\
NGC 1900$^*$    &	79.7875	&	-63.0240    & -0.55&0.05  & -0.12   & 0.011   & 8.842    & 0.01     &  1        & 1  \\
NGC 1978	    &	82.1889	&	-66.2367    & -0.49&0.09  & -0.097  & 0.003   & 9.4      & 0.12     &  2        & 4  \\
NGC 2005$^*$	&	82.5417	&	-69.7530    & -1.75&0.08  & -0.085  & 0.003   & 10.12    & 0.04     &  2        & 3  \\
NGC 2019$^*$	&	82.9833	&	-70.1590    & -1.41&0.06  & -0.074  & 0.004   & 10.1     & 0.05     &  3        & 3  \\
NGC 2121$^*$	&	87.0500	&	-71.4810    & -0.54&0.06  & -0.087  & 0.003   & 9.246    & 0.23     &  1        & 4  \\
NGC 2155$^*$	&	89.6375	&	-65.4770    & -0.59&0.04  & -0.13   & 0.003   & 9.45     & 0.07     &  2        & 4  \\
NGC 2210$^*$	&	92.8792	&	-69.1210    & -1.74&0.11  & -0.086  & 0.003   & 10.08    & 0.04     &  2        & 3  \\
NGC 2249$^*$	&	96.4542	&	-68.9200    & -0.45&0.12  & -0.091  & 0.003   & 9.08     & 0.1      &  2        & 11 \\
NGC 339$^*$     &	14.4484	&	-74.4707    & -1.15&0.09  & -0.093  & 0.002   & 9.77     & 0.17     &  2        & 2  \\
NGC 361$^*$    	&	15.5463	&	-71.6070    & -0.9 &0.03  & -0.103  & 0.003   & 9.51     & 0.1      &  6        & 2  \\
NGC 416	        &	16.9960	&	-72.3557    & -0.85&0.08  & -0.098  & 0.005   & 9.78     & 0.08     &  2        & 2  \\
NGC 419	        &	17.0719	&	-72.8838    & -0.62&0.2   & -0.099  & 0.004   & 9.3      & 0.13     &  2        & 2  \\
SL 388$^*$	    &	80.0208	&	-63.4800    & -0.39&0.09  & -0.067  & 0.004   & 9.002    & 0.12     &  1        & 12 \\
SL 509$^*$	    &	82.4500	&	-63.6490    & -0.54&0.05  & -0.083  & 0.004   & 9.138    & 0.18     &  1        & 1  \\
SL 817$^*$      &	90.1583	&	-70.0690    & -0.41&0.03  & -0.067  & 0.008   & 8.877    & 0.08     &  1        & 1  \\
SL 862$^*$	    &	93.3625	&	-70.6960    & -0.47&0.1   & -0.078  & 0.011   & 8.851    & 0.09     &  1        & 1  \\
\hline
\end{tabular}
%\footnote{
\end{center}
%References:-
%
$^*$: Observed clusters\\
Age:
1: \cite{dhanush2024}
2: \cite{milone2023}    
3: \cite{narloch2022}
4: \cite{gatto2021}   
5: \cite{parisi2022}
6: \cite{narloch2021} 
[Fe/H]:
%0.	...
1.  \cite{sharma2010}
2.  \cite{parisi2022}
3:  \cite{mucciarelli2021}  
4:  \cite{song2021}
5.  \cite{beasley2002}
6.  \cite{milone2023}    
7.  \cite{grocholski2006}
8.  \cite{dacosta1}
9.  \cite{parisi2014}
10. \cite{narloch2022}
11. \cite{kerber2007}
12. \cite{brocato2001}
\label{param}
\end{table*}

\section{Observations and data reductions}

Most observations were obtained over three   continuous
nights on December 19-21, 2004, with the WFIRC on the 2.5-m du Pont telescope
at the Las Campanas observatory (see \cite{persson} for details). The WFIRC
(Wide Field Infrared Camera) dewar houses four Rockwell 1024 x 1024 pixel
HgCdTe arrays. The pixel size is 18.5 microns, which gives a scale of
0.196 arcsec/pixel and a linear dimension of $201\times201$ arcsec for each
array at the Cassegrain focus of the 100-inch telescope. The chip No. 2
has the betabnew.dat.smcst cosmetics and was used for the observations of the
clusters. Chips No. 1,3,4 were used for the background subtraction
of the ``field'' stars.

After the standard reductions of the IR frames, the stellar photometry was
carried out with IRAF DAOPHOT II  (\cite[see also,][]{sharma2020, sharma2016}).
%(\cite{stetson}). 
The final photometry list contains $J$
and $K_S$ magnitudes of 17665 stars with ``formal'' DAOPHOT errors less
than $0.15$ mag in each filter. The limiting magnitudes of our photometry are: $J_{\rm lim}
= 18.3 - 19.8$ and $K_{\rm S\,lim} = 17.5 - 19.0$. Between $50$ and
$700$ 2MASS stars in each field were used to obtain the zero points for
transformation to the standard $JHK_S$ system. 
%No color terms were included.

The observations of the clusters ESO121-3, NGC 2121,
NGC 2249,  SL 388,  SL 509,  SL 817, and SL 862 are presented in this paper
were collected with the ESO/NTT telescope at the La Silla observatory
in Chile, in the course of the Araucaria project (\cite{pietrzynski}) with
the main goal is to study the properties of the red clump (RC) stars in the IR domain
(e.g. \cite{pietrzynski1}). These clusters were observed with the SOFI instrument
in wide-field mode with a focal elongator in the grism wheel (LFFO)
during three different photometric nights together with a large
number  (typically 6-8 per night) of standard stars from the UKIRT
list (\cite{hawarden}). The resulting field of view was about
$2.49\times 2.49$  arcmin square with a scale of 0.146 arcsec/pixel.
Since 2005, the LFFO SOFI setup was not offered by ESO. Hence, SOFI observed another three clusters (Lindsay 1, NGC339, Lindsay 113) with a different wide-field mode.
In this configuration, the field of view
was $4.92\times4.92$ {arcmin square with a scale of 0.288 arcsec/pixel.
The total integration times  were about 20 and 3 minutes in the
$K_S$ and $J$ band filter, respectively. The relatively loose clusters
were observed using the so-called simple jittering technique, while in the
case of the dense ones, the sky was probed a few arcmin away
from the cluster. The PSF photometry was obtained with the DAOPHOT and
ALLSTAR programs as described in detail in \cite{pietrzynski2}. The
instrumental magnitudes were transformed to the 2MASS photometric system in the same way as 
for du Pont telescope observations.

\section {Result and analysis}

\subsection{The RGB Slope}

The RGB slope is related to the effective temperature of the stars along the RGB, 
and $T_{eff}$ in turn depends on the opacity and heavy element abundance.
This parameter is very powerful because it is independent of reddening and distance. It is
significantly less demanding in terms of observing time and telescope collecting area than the spectroscopic methods \cite{ivanov2002}.
 \cite{ivanov2002} and \cite{Kuchinski1995} argued that the RGB morphology can be very well constrained through the RGB slope, which is linearly fitted between 0.6 and 5.1 mag brighter than the zero-age horizontal branch (ZAHB).
However, \cite{valentia} argued that in the case of low-intermediate-metallicity clusters,
the accurate measurement of the location of the ZAHB in IR CMD is impossible
because the HB is not horizontal at all. To apply a homogeneous procedure to the entire cluster
sample, they fit the RGB in the magnitude range between 0.5 and 5 mags fainter than the
brightest star of each cluster after previous decontamination by the asymptotic giant branch and field stars.
Also, \cite{Kuinskas2008} studied 14 LMC/SMC clusters and derived RGB slopes 
using the slightly different criteria, using RGB stars in the range $M_{Ks}$ = $-2.0$ to  $-6.4$ for fitting. 
\cite{kyeong} in their study of RGB slopes from old open Galactic clusters, include 
those stars in the giant branch only with absolute magnitude in the range $-1.5 \leq M_{K} \leq -6.5$. 
This criteria is similar to \cite{tiede}'s
approach. This criterion excludes both RC stars and bright asymptotic giant branch stars.

In this study, we followed the procedure to determine the RGB slope
described in \cite{ivanov2002}, carrying out a least-squares fit to the RGB from
0.5 mag above the horizontal branch (cf. Fig. \ref{cmd}) using the equation:
\begin{equation}
(J-K_S) = {\rm ZP + Slope_{RGB}} \times K_S
\end{equation}
on the statistically cleaned CMD to derive the slope and the zero point,
assuming $K_S$ to be the independent variable.
The LMC and SMC are relatively nearby targets, and the
globular clusters span a wide range of metallicity and age. 
From 23 clusters of our sample, there are 4 SMC clusters (Lindsay\,1, Lindsay 113, NGC\,339, 361)
and the rest 19  belong to LMC. From our photometry, we got good quality NIR data (error $<$ 0.1 for $K$=18 mag), and it has been used to generate CMDs for all the target clusters.
 Cluster membership criteria is crucial in defining the cluster parameters.
As we can't reach the faint limits of our NIR observation through Gaia's proper motion data, 
we statistically subtracted the contribution of field stars from the CMD of the cluster region
following the approach of \cite{sharma2017}.
The CMD of the cluster and field region were divided into rectangular bins of size 0.5 in $K_S$ mag and 0.25 mag in $J-K_S$ color.
Then, from each bin of the cluster CMD, we randomly removed a number of
stars equal to the number of stars in the corresponding bin of the ``field'' CMD.
We have defined the cluster region as a circular area corresponding to the angular
dimension of the cluster where the stellar density reaches the background density.
The cluster centers and radii are chosen by eye, based primarily on the photometric catalog. 
We have used a circular annulus around the cluster having the same area as the cluster region for a field star distribution.  
Fig. \ref{cmd} shows $K, (J-K)$ some examples of CMDs for the stars in clusters regions (left panels) and corresponding field regions (middle panels). We show clusters with different ages and metallicities.
The contamination due to background field population is visible in the cluster region CMDs, which, in some cases, is $\sim$40-50\% of the stars in the designated cluster area.
In most of the older clusters, these two populations are almost completely overlaid and only precise
statistical analysis can distinguish them.
The statistically cleaned $K, (J-K)$ CMD of the cluster region
is also shown in the right panels of Fig. \ref{cmd}, clearly showing the distribution of RGB stars in the clusters.
The RGB locus was then defined after inspecting each CMD. We removed
10-$\sigma$ outliers, and repeated the fitting. Typically,
the fitting coefficients obtained after two iterations were
statistically indistinguishable. The probable cluster member stars, which are used to calculate the slope of RGB, are shown in red circles in Fig. \ref{cmd} for a few sample clusters. A total of 300 realizations
were generated using Monte Carlo simulations for each cluster -- each including the contamination removal and the fitting -- and we averaged the fitting results (see also
\cite{ivanov2002} and \cite{ivanov2000}). The derived value of the slope of the RGB stars of the clusters is given in Table \ref{param}.

For four clusters in our sample, namely NGC\,1900, NGC\,2155, NGC\,2005 and NGC\,1788  we noticed some discrepancy of the derived RGB slopes from the main trend of the rest of the clusters. We performed additional checks to validate the slope value (or not). Using the VISTA Science Archive (VSA, http://vsa.roe.ac.uk/index.html), we retrieved the merged  $J$ and $Ks$ photometric catalogs  ($5\times5$ arcmin square field around the center of each cluster) from Vista Magellanic Cloud NIR survey (VMC, \cite{cioni2011}). Then, the stars are cross-matched with the Gaia EDR3 proper motion and photometry catalog (\cite{gaia2016}). In addition to the statistical cleaning used above, the most probable cluster members are verified by the combination of proper motions and VMC  $Ks, (J-Ks)$ color-magnitude diagrams. 
The process is illustrated in Fig. \ref{RGBcheck}, and we got similar RGB slope values.

\begin{figure*}
\centering\includegraphics[width=1.1\textwidth]{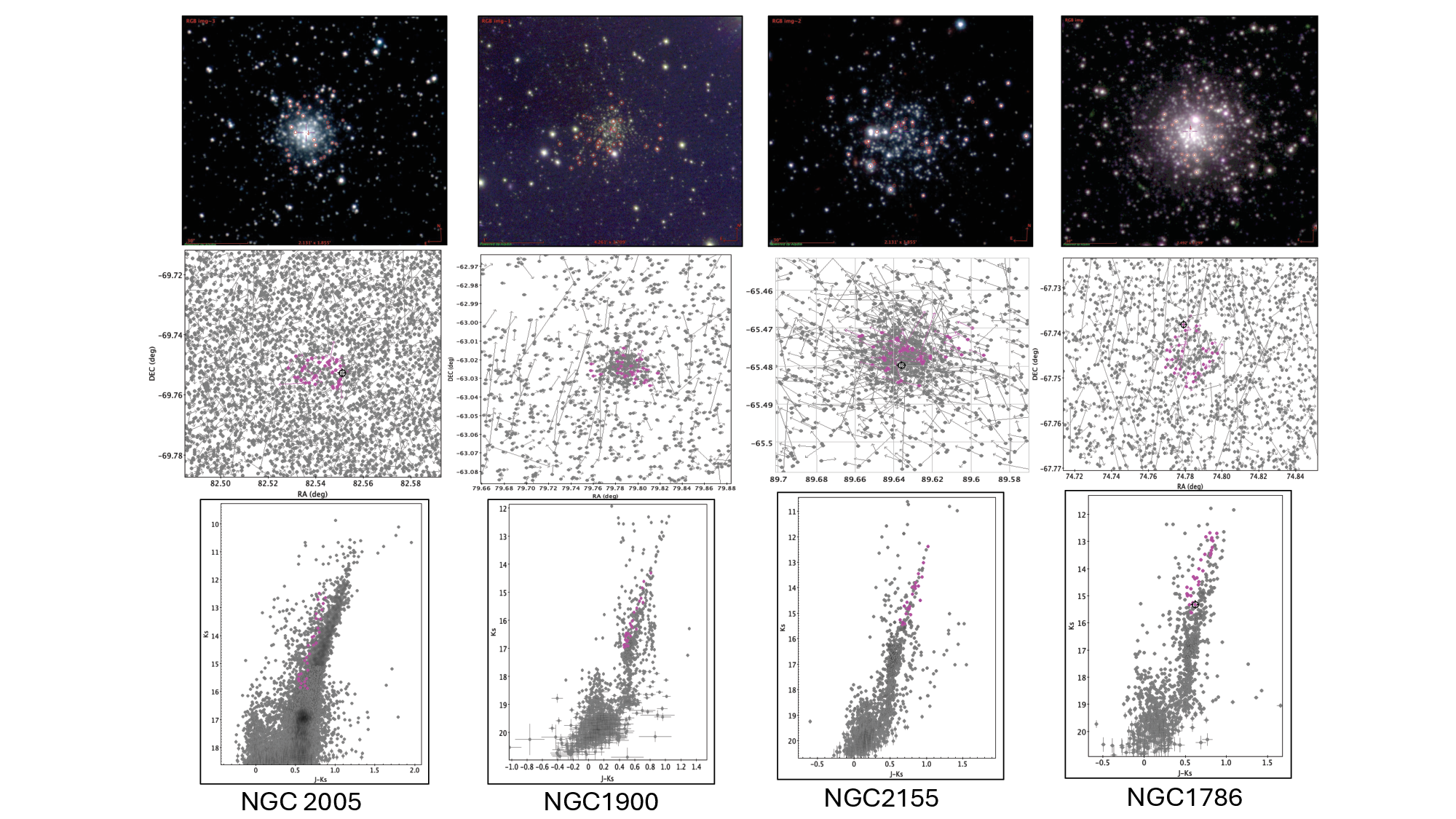}
\caption{\label{RGBcheck} The false color  $Y$, $J$ and $Ks$ images, proper motion diagrams and color-magnitude diagrams of NGC\,1900, NGC\,2155, NGC\,2005 and NGC\,1788. The images are taken from the Vista Magellanic Cloud survey. The black dots are all stars in the $5\times5$ arcmin square field around the center of each cluster. The most probable RGB members used to derive the RGB slope are marked with red circles.
}
\end{figure*}

\subsection{RGB slope-metallicity-age calibration}

A sample of star clusters is constructed as follows: for LMC and SMC, we used our data
presented in this paper plus \cite{Kuinskas2008} sample transformed to our RGB slope estimation.
The ages of the LMC and SMC clusters are taken from the literature, using mainly two recent works \cite{dhanush2024} and \cite{narloch2022}, to maintain the homogeneity.  They are listed in the Table \ref{param} with the corresponding references. Our sample contains both young and old clusters, the youngest cluster in the sample is NGC\,1900, with an age of 700 Myr, and the oldest one NGC\,2005 has an age of around 13 Gyr.  The metallicities of the clusters in our sample are based on the latest and most reliable measurements in the literature. They are also listed in Table \ref{param} with the corresponding references.   
 
As a consistency check of the homogeneity of the age and metallicities taken from the literature of such constructed star cluster sample, we plotted the well-known Age-metallicity relation (AMR) in the right panel of Fig. \ref{age_feh}. The red circles represent SMC clusters, and the blue circles represent the LMC sample. In the left panel, we compare the AMR relation for the LMC and SMC clusters sample taken from  \cite{narloch2022}, their Fig.13. 

\begin{figure*}
\centering
\includegraphics[width=0.47\textwidth]{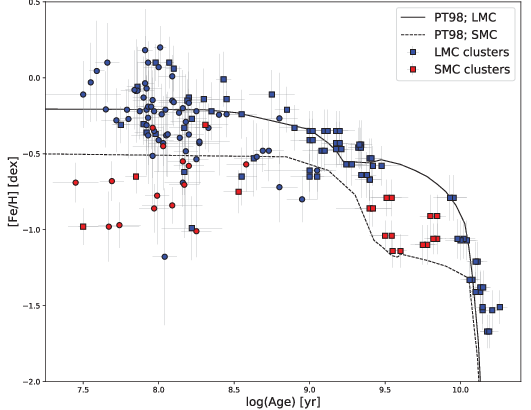}
  \includegraphics[width=0.50\textwidth]{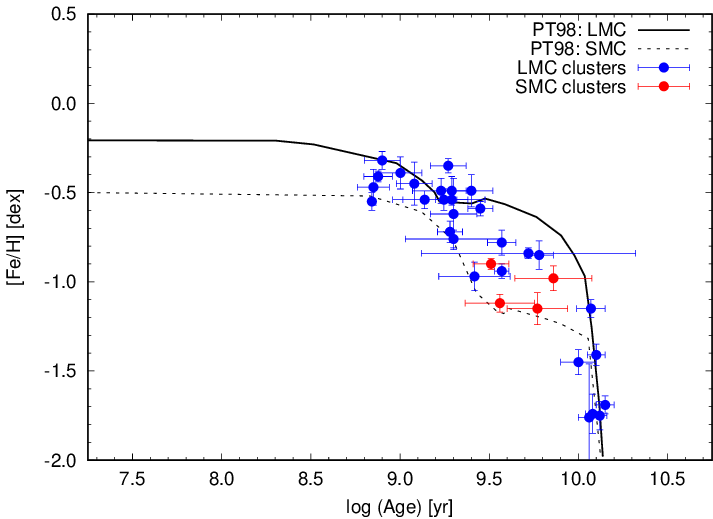}
\caption{Left panel: Log(Age) vs. [Fe/H] (AMR) relation for the LMC and SMC clusters sample taken from \cite{narloch2022}, their Fig.13. Right panel: Log(Age) vs. [Fe/H] (AMR) relation for the present LMC and SMC clusters sample. Blue circles represent LMC clusters, and the red circles represent the SMC sample. The solid and dashed lines correspond to LMC and SMC from the bursting models of \cite{pagel1998}.
}
\label{age_feh}
\end{figure*}

The LMC sample follows the bursting models of \cite{pagel1998}, with the age gap between 3 and 10 Gyr. The low-metallicity clusters are situated before the age gap, while the most metal-rich clusters have been formed recently (\cite[for more details,  see][]{sharma2010}). The SMC clusters are too few to make any conclusion.

 The next relation we build in our analysis is the log (Age) - RGB slope for all clusters in our sample, shown in Fig. \ref{age_slope}. As seen from the plot, there are no shifts in the location on the diagram of the LMC and SMC clusters; therefore, they can be used as homogeneous samples. The RGB slope do not shows any statistically significant age dependence. The correlation coefficient of both groups (LMC and SMC) is around 0.05.

\begin{figure}
\centering
\includegraphics[width=0.50\textwidth]{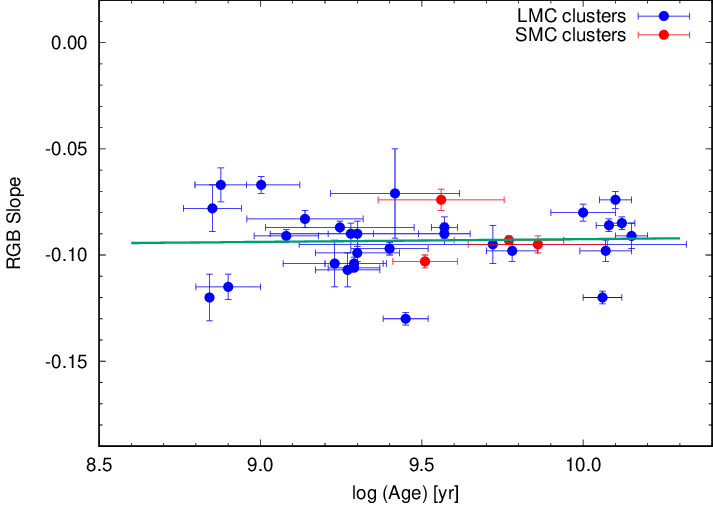}
 \caption{The Log (Age) vs. RGB slope relation for the clusters from our sample. The symbols are the same as in Fig. \ref{age_feh}.  The solid line is the least square fit to the whole cluster sample.} 
\label{age_slope}
\end{figure}

And finally, we have plotted the RGB slope as a function of [Fe/H] for all clusters in our sample (Fig. \ref{slope_feh}). Magenta circles represent the whole sample, the black ones for clusters younger than 3\,Gyr, and the grey circles stand for clusters older than 10\,Gyr.

\begin{figure}
\centering
\includegraphics[width=0.49\textwidth]{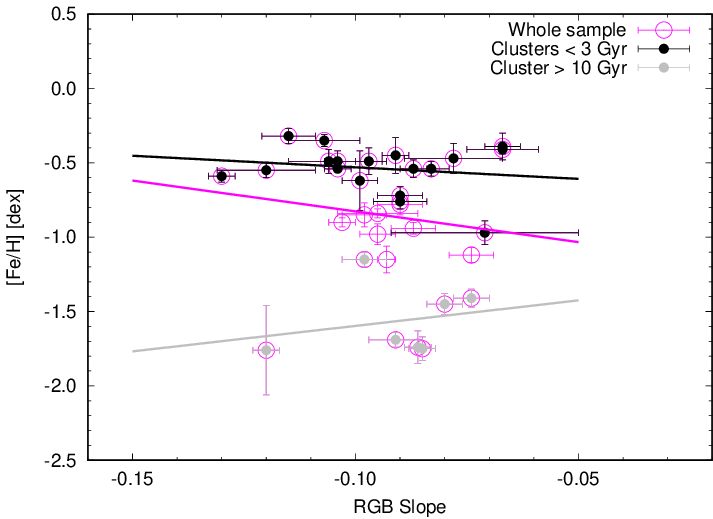}
\caption{The RGB slope vs. [Fe/H] relation. The black circles represent the clusters younger than 3\,Gyr, and the grey ones represent clusters older than 10\,Gyr. Magenta circles represent the whole cluster sample. The solid lines represent the derived best fit for the corresponding sample.
}
\label{slope_feh}
\end{figure}

The least-square fit to the whole sample is:
\begin{eqnarray}
\rm {\rm [Fe/H]  =-4.13_{\pm 1.2} \times  Slope_{JK}^{RGB}}-1.24_{\pm 0.5}
\end{eqnarray}

The least-square fit to the clusters older than 10\,Gyr is:

\begin{eqnarray}
%\rm {\rm [Fe/H]  =-6.27_{\pm 0.9} \times  Slope_{JK}^{RGB}}-1.72_{\pm 0.22}
\rm {\rm [Fe/H]  =2.64_{\pm 0.9} \times  Slope_{JK}^{RGB}}-1.33_{\pm 0.22}
\end{eqnarray}

The least-square fit to the clusters younger than 3\,Gyr is:

\begin{eqnarray}
%\rm {\rm [Fe/H]  =-0.067_{\pm 0.5} \times  Slope_{JK}^{RGB}}-0.52_{\pm 0.8}
\rm {\rm [Fe/H]  =-1.55_{\pm 0.5} \times  Slope_{JK}^{RGB}}-0.69_{\pm 0.8}
\end{eqnarray}

The conservative uncertainty of the individual photometric cluster metallicity estimate, due to uncertainties in the
cluster slope derivation, age, metallicity, etc., can be estimated as 0.3 dex.

As can be seen, the young clusters do not show statistically significant  RGB slope [Fe/H] dependence, and some weak relations can be found for the older clusters.

\section{Conclusions}

This paper presents deep near-IR $JHK$ stellar photometry for 23 LMC/SMC globular clusters. SOFI and Du-pont images enable us to go as deep as 18 mags in the K band; therefore, for the majority of the clusters,  the $J-K_S, K_S$ CMDs reveal well-defined RGBs.
The well-defined RGBs of the cluster have been used to determine their slopes which in turn are used 
to determine the relation between slope-age-metallicity.
We checked the AMR for the present sample and found LMC clusters follow the burst model of \cite{pagel1999}, with an age gap between 3-10 Gyr.
The RGB slope of the present LMC/SMC cluster sample does not show any statistically significant age dependence. 
We have checked the dependence of the RGB slope on the metallicity for all 33 LMC/SMC clusters in the present sample.
They follow the prediction of the theoretical models. The young and old clusters are found to be distributed differently in RGB slope-metallicity space. The population of the younger clusters shows a marginal negative slope, whereas the older clusters show a positive slope.

\backmatter

%\bmhead{Supplementary information}

%If your article has accompanying supplementary file/s please state so here. 

%Authors reporting data from electrophoretic gels and blots should supply the full unprocessed scans for key as part of their Supplementary information. This may be requested by the editorial team/s if it is missing.

%Please refer to Journal-level guidance for any specific requirements.

\bmhead{Acknowledgements}

The data used in this paper have been obtained with du Pont telescope at Las Campanas observatory and ESO NTT at La Silla observatory. 
Support for SS and  JB are also provided by ANID’s Millennium Science Initiative ICN12-009, awarded to the Millennium Institute of Astrophysics (MAS).

\section*{Declarations}

%Some journals require declarations to be submitted in a standardised format. Please check the Instructions for Authors of the journal to which you are submitting to see if you need to complete this section. If yes, your manuscript must contain the following sections under the heading `Declarations':

\begin{itemize}
%\item Conflict of interest/Competing interests (check journal-specific guidelines for which heading to use)
%\item Ethics approval and consent to participate
%\item Consent for publication
\item Data availability: 
 Fig.1: Available on request.
 Fig 2, Fig 3, Fig 4: Available in Table 1.
\end{itemize}

\end{document}